\documentclass[aps,prl,twocolumn,showpacs,superscriptaddress,groupedaddress]{revtex4}

\usepackage[T1]{fontenc}
\usepackage[utf8]{inputenc}
\usepackage[english]{babel}
\usepackage{bm}	%For bold math symbols
\usepackage{amsopn}	%For DeclareMathOperator command
\usepackage{amssymb}	%For hollow letters (real and complex numbers)
\usepackage{amsmath}	%For intext \smallmatrix
\usepackage{dsfont}		%For identity operator
\usepackage{braket}		%For easy bra and ket
\usepackage{physics}
\usepackage{graphicx}
\usepackage{svg}
\usepackage{booktabs}	%For tabular
\usepackage{makecell}	%For multi-line table cells
\usepackage{tabularx}	    %For tabularx

\usepackage{tikz}
\usetikzlibrary{arrows,positioning}
\usepackage{environ}
\usepackage{varwidth}
\begin{document}

% Use the \preprint command to place your local institutional report
% number in the upper righthand corner of the title page in preprint mode.
% Multiple \preprint commands are allowed.
% Use the 'preprintnumbers' class option to override journal defaults
% to display numbers if necessary
%\preprint{}

%Title of paper
\title{Position- and momentum-squeezed quantum states in micro-scale mechanical resonators}

\author{Yann Le Coq}
\affiliation{LNE-SYRTE, Observatoire de Paris, Universit\' e PSL, CNRS, Sorbonne Universit\' e, Paris, France}
\author{Klaus M{\o}lmer}
\affiliation{Department of Physics and Astronomy, Aarhus University, Ny Munkegade 120, DK-8000 Aarhus C, Denmark}
\author{Signe Seidelin}
\email{signe.seidelin@neel.cnrs.fr}
\affiliation{Univ. Grenoble Alpes, CNRS, Grenoble INP and Institut N\' eel, 38000 Grenoble, France}
\affiliation{Institut Universitaire de France, 103 Boulevard Saint-Michel, 75005 Paris, France}

\date{\today}

\begin{abstract}
A challenge of modern physics is to investigate the quantum behavior of a bulk material object, for instance a mechanical oscillator. We have earlier demonstrated that by coupling a mechanical oscillator to the energy levels of embedded rare-earth ion dopants, it is possible to prepare such a resonator in a low phonon number state. Here, we describe how to extend this protocol in order to prepare momentum- and position squeezed states, and we analyze how the obtainable degree of squeezing depends on the initial conditions and on the coupling of the oscillator to its thermal environment.
\end{abstract}
% insert suggested PACS numbers in braces on next line
\maketitle

 \section{Introduction}

Research in modern physics has led to impressive progress in the field of quantum physics, quantum materials, and quantum engineering. The study of the quantum behavior of mechanical, macroscopic systems, is of foundational importance and it plays important roles in sensing and in transducers, where mechanical oscillators may bridge the coupling among optical, microwave, and acoustic waves and discrete quantum degrees of freedom. One major difficulty concerning mechanical, macroscopic systems relies in interacting with their motion without destroying their quantum behavior. One approach consists of exploiting a hybrid quantum system consisting of a mechanical oscillator coupled to an atom-like object, and interact via the atom-like object for which the quantum nature is less fragile, and for which the interrogation methods have been perfectioned over the course of several decades. Using this hybrid approach, a research team managed in 2011 to place a mechanical high-frequency resonator in the quantum ground state, by coupling it to a superconducting quantum bit \cite{oconnell2010}. This achievement has been a major driving force in the expansion of the field of mechanical quantum hybrid systems, and since then, many alternative systems have been investigated both theoretically and experimentally \cite{treutlein2014}.

\section{Physical system}

We have previously proposed a quantum hybrid system consisting of a crystalline mechanical resonator which contains rare-earth ion dopants \cite{molmer2016}, and we have recently demonstrated how such resonator can be cooled significantly below the surrounding thermal bath temperature \cite{seidelin2019}. We will in the following summarize the main ideas of these schemes, as they constitute the starting point for the work presented here.

In the general case of a bulk crystal resonator, the oscillations generate a mechanical strain, and this strain influences the electronic properties of the impurities or dopants, as a consequence of the modification of their local environment, including the electronic orbital distributions. The oscillations of the bulk resonator are thus mapped onto the energy levels of the impurity. This type of ``strain-coupling'' has previously been observed with a single quantum dot in an oscillating photonic nanowire \cite{Yeo2014}, and with a single nitrogen vacancy in diamond \cite{Teissier2014,Ovartchaiyapong2014}.

In our case, we wish to benefit from the record-long coherence times of rare-earth dopants, in particular europium ions in an $\rm Y_2SiO_5$ crystal matrix. Due to the difficulty of detecting single europium ions, we will instead focus this proposal on ion ensembles. In order to avoid effects due to inhomogeneous broadening, we suggest to use the technique of spectral hole burning, in which a fraction of the ions at a given energy is selectively pumped to a dark state, leaving a transparent spectral window in the inhomogeneous absorption line behind. In brief, our strain-coupling protocol \cite{molmer2016} is based on the fact that when the resonator oscillates, the strain periodically broadens and narrows the spectral hole (it ``breathes''). If we use a probe laser beam that is close in frequency to the edge of the spectral hole, but within the transparent window, it will couple dispersively to the edge of the spectral hole. This interaction leads to a modulation of the phase of the transmitted laser, and this phaseshift can be used to monitor the position and momentum of the resonator. Moreover, the laser beam shifts the equilibrium position of the resonator, and by using an active feedback protocol it can cool the resonator \cite{seidelin2019}. In addition, as we will show in this article, the laser beam can be applied in a certain temporally modulated fashion in order to create a position- or moment squeezed mechanical resonators.

\begin{figure}[t]
\centerline{\includegraphics[width=9cm]{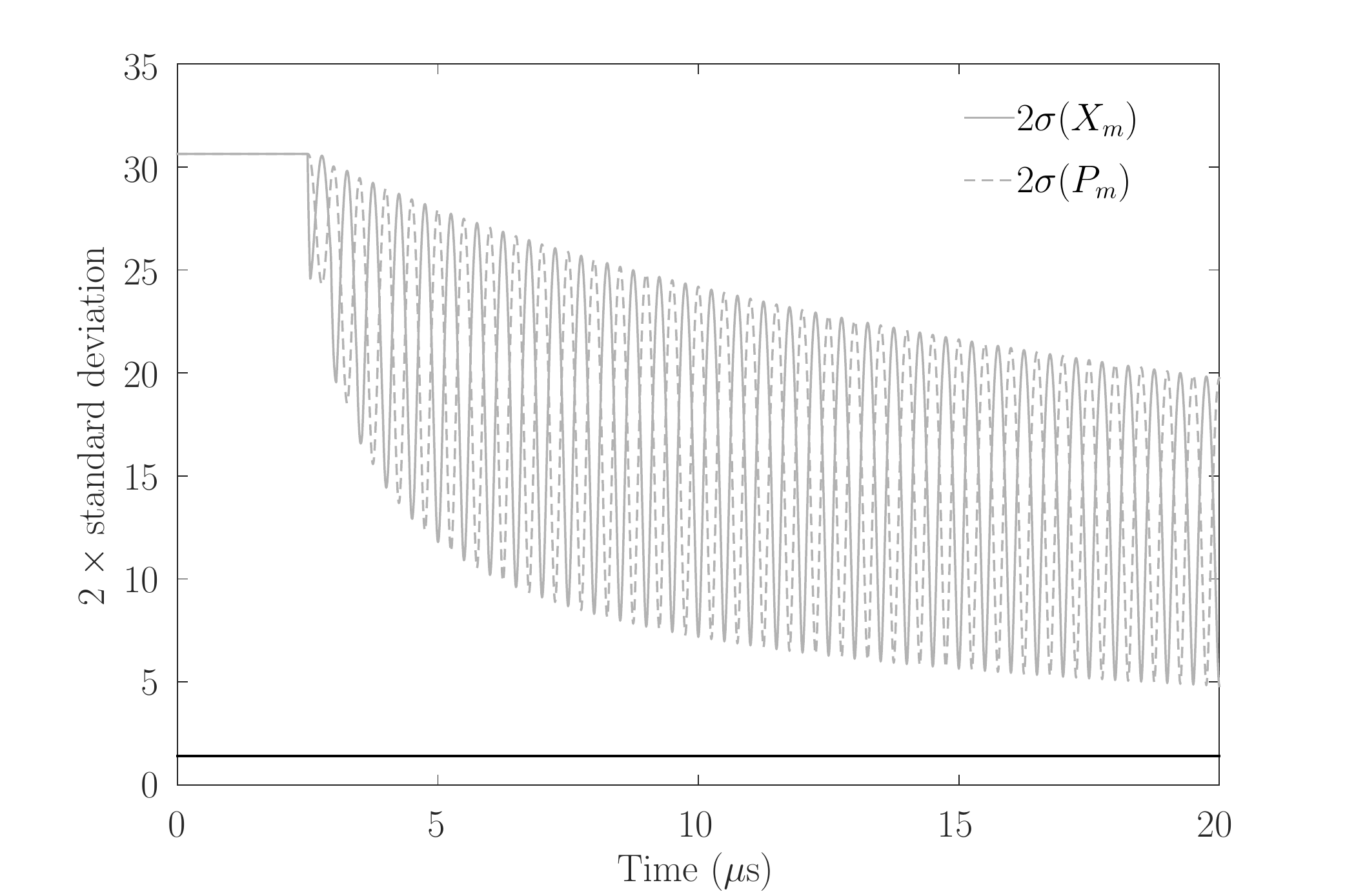}}
\vspace*{8pt}
\caption{The first 20 microseconds of the squeezing sequence with a bath temperature of 10 mK, allowing to visualize the individual oscillations of two times the standard deviation of $X_m$ and $P_m$. The initial flat part of the curve (lasting 2.5 mechanical periods of oscillation) corresponds to the time prior to application of the squeezing pulses, in order to allow the visualization of  the steady state. The solid straight line at the bottom of the graph at 1.41 corresponds to the value below which $X_m$ or $P_m$ must drop in order to attain squeezing. The full duration of the simulation is shown in figure~\ref{10mK}.\label{zoom_10mK}}
\end{figure}

\section{Squeezing by stroboscopic probing}

Using a Gaussian ansatz developed in earlier publications \cite{Adesso2014,seidelin2019} allows one to establish the equations of motion for the variances of the position and momentum ($a_{11}/2$ and $a_{22}/2$ respectively) as well as the co-variances ($a_{12}/2$ and $a_{21}/2$) of the resonator:

\begin{eqnarray}
\frac{da_{11}}{dt} & = & -\eta \kappa^2 a_{11}^2 + \omega(a_{21}+a_{12})-\gamma(a_{11}-(2\overline{n}+1))\nonumber \\
\frac{da_{12}}{dt} & = & - \eta \kappa^2 a_{11} a_{12} - \omega(a_{11}-a_{22})-\gamma a_{12}\nonumber \\
\frac{da_{21}}{dt} & = & - \eta \kappa^2 a_{11} a_{21} - \omega(a_{11}-a_{22})-\gamma a_{21}\nonumber \\
\frac{da_{22}}{dt} & = & \kappa^2 - \eta \kappa^2 a_{12} a_{21} - \omega(a_{21}+a_{12}) -\gamma(a_{22}-(2\overline{n}+1)).\nonumber\\
\label{diffeq}
\end{eqnarray}

Here, $\eta$ corresponds to the measurement efficiency, $\omega$ is the mechanical oscillator frequency, and $\gamma$ the rate of coupling between the resonator and the environment. The constant $\kappa$ reflects the interaction strength between the laser beam and the resonator which depends on photon flux as well as the strain sensitivity. More precisely, by interacting with a bent cantilever with an appropriately prepared spectral hole structure, the light beam experiences a phase shift $\Delta\phi$ proportional to the dimensionless resonator displacement $X_{m}=x_m/x_0$ (with $x_0=\sqrt{\hbar/m\omega}$), namely $\Delta\phi = \beta X_{m}$, with a  proportionality constant $\beta$, that depends on the strain sensitivity, the resonator geometry, as well as the protocol employed for the spectral hole burning \cite{molmer2016, seidelin2019}. The constant $\beta$ is linked to $\kappa$ in the following way: $\kappa^2\equiv 2 \beta^2\Phi$, where $\Phi$ is the photon flux of the probe laser.\\

\begin{figure}[t]
\centerline{\includegraphics[width=9cm]{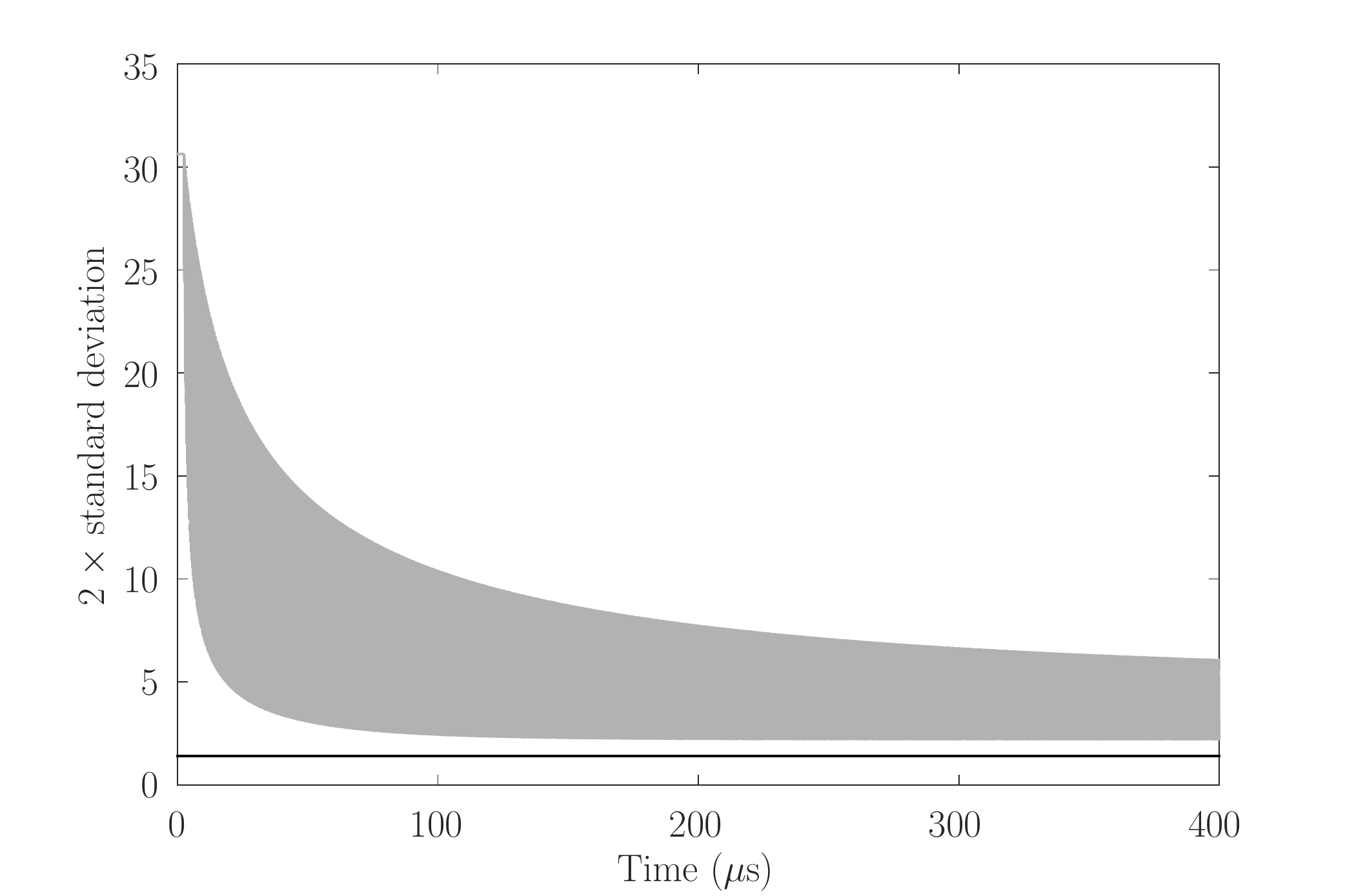}}
\vspace*{8pt}
\caption{Same parameters as in figure \ref{zoom_10mK} but extending the timescale out to 400 $\mu$s. Also here, the curves show two times the standard deviation of $X_m$ and $P_m$, but the oscillations are so fast that they cannot be distinguished and appear as a gray surface. The minimum value of both variables never drops below 1.41 (solid line), which indicates the absence of squeezing.\label{10mK}}.
\end{figure}

To achieve squeezing, we apply the probe laser stroboscopically \cite{Thorne1978,Braginsky1980,Vasilakis2015,Wade2015}. That is, we pulse the probe laser such that it is on only for a fraction of the mechanical oscillation period,  typically for approximately one tenth of the oscillation period for each pulse. This amounts to having $\kappa$ equal to zero in Eqs.~\ref{diffeq} at all times except for short periodic time intervals. The laser pulse has the effect of measuring the resonator position $X_m$ which squeezes this degree of freedom (that is, decreases its uncertainty) and antisqueezes (increases uncertainty) the other quadrature $P_m$. These effects are represented by the $-\eta\kappa^2 a_{11}$ and $+\kappa^2$ terms in the first and last line of Eq.(1), respectively. As the ellipse representing the gaussian ansatz is continuously rotating in phase space (terms proportional to $\omega$ in Eqs.~\ref{diffeq}), this stroboscopic measurement protocol progressively reduces the size of the ellipse toward the Heisenberg's limit, and once this limit is sufficiently approached, it induces quantum squeezing and antisqueezing of the quadratures. On the other hand, thermal coupling to the non-zero temperature environment tends to increase the ellipse's size away from the Heisenberg's limit and toward the thermal equilibrium limit (terms proportional to $\gamma$ in Eqs.~\ref{diffeq}). The degree of squeezing that can finally be obtained depends on the physical parameters of the setup, in particular the environment temperature and the coupling rate between this environment and the resonator, as we will discuss in the following.

\section{Result of simulations} 

\begin{figure}[t]
\centerline{\includegraphics[width=9cm]{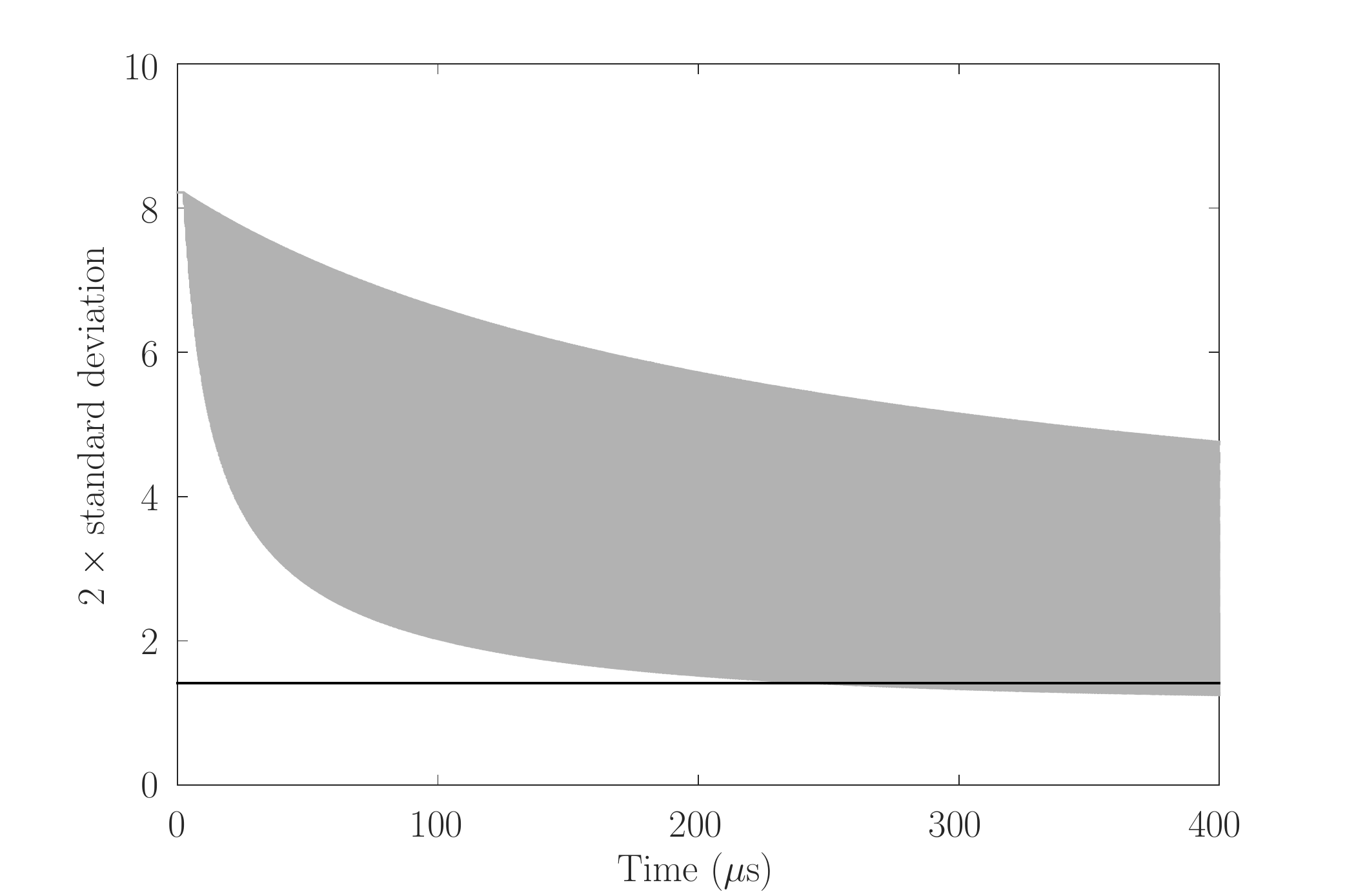}}
\vspace*{8pt}
\caption{Simulations of squeezed states with a bath temperature of 0.7 mK. We observe a modest squeezing effect (the gray area drops below the solid line) from approximately 250 $\mu$s and forward.\label{0_7mK}}
\end{figure}

We solve Eqs.~\ref{diffeq} with realistic physical parameters, based on an existing experimental setup described in refs.~\cite{Gobron2017,Motte2019}. We use the example of a $100\times10\times10$\,$\mu m^3$ cantilever with a bending mode frequency of $\omega =2\pi\times 1$\,MHz mode and an effective mass $m=1.1\cdot 10^{-11}$\,kg. The resonator material consists of $\rm Y_2SiO_5$ containing a  0.1 \% doping of $\rm Eu^{3+}$ ions, with a $^7F_0 \rightarrow$  $^5D_0$ transition centered at 580 nm and a measured linewidth of approximately $2\pi \times1$\,kHz \cite{Gobron2017}. We use the strain sensitivity of the $\rm Eu^{3+}$ ions of -211.4 Hz/Pa~\cite{Thorpe2011}, and a spectral hole width of 6~MHz. Moreover, we assume a laser intensity of 1\,mW (\emph{i.e.} a photon flux of $\Phi=2.92 \times 10^{15} s^{^-1}$), and the coupling to the environment of $\gamma = 2\pi\times 10$\,Hz (if nothing else is stated). Furthermore, we assume $\eta=1$, and we will investigate the degree of squeezing as a function of the bath temperature, assuming that the resonator is at thermal equilibrium with the bath when we start the sequence of squeezing pulses. In order to do the stroboscopic measurements we modulate the interaction strength $\kappa$. In practice, we solve Eqs.~\ref{diffeq}, with a non-zero interaction strength $\kappa$ only as long as $\vert \cos(\omega t)\vert>0.9$ and $\kappa=0$ for all other times $t$. Concerning the numerical value of $\kappa$, we previously determined \cite{molmer2016} that a displacement of the resonator tip equal to 0.4 pm gives rise to a phase shift $\Delta \phi$ of 0.2 mrad such that $\beta=\Delta \phi/X_m=(\Delta \phi/x_m)x_0 =(0.2 \times 10^{-3}/0.4\times 10^{-12})1.3\times10^{-15}=0.65 ~\mu\rm{rad}$, such that $\kappa^2= 2 \beta^2\Phi=2\pi\times 197$\,Hz (this value corresponds to an average value over the full mechanical oscillation; during a stroboscopic pulse, the peak value is ten times higher).

\begin{figure}[t]
\centerline{\includegraphics[width=9cm]{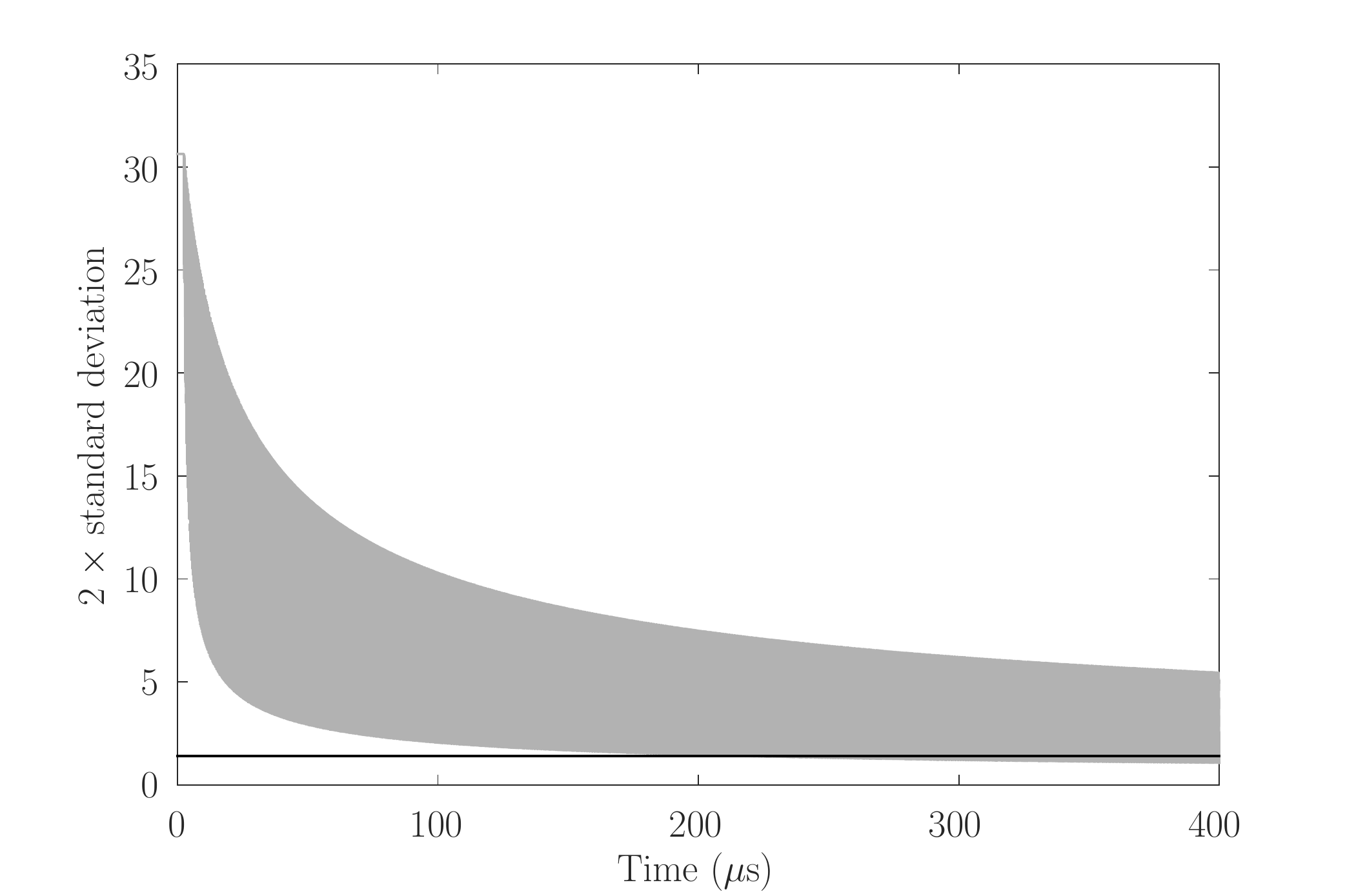}}
\vspace*{8pt}
\caption{Here, the temperature of the environment is again set to 10 mK, but we have reduced the rate of coupling to the environment to $\gamma$=0.1 Hz, which also allows one to obtain a weak effect of squeezing.\label{10mk_gamma_0_1}}
\end{figure}

We first investigate the case in which the bath temperature is 10 mK; the results are shown in figure~\ref{zoom_10mK} and~\ref{10mK}. In these, and in the following figures, we plot two times the standard deviation of the resonator position and momentum, $\sigma (X_m)$ and $\sigma (P_m)$, where $P_m$ is also given in dimensionless units: $P_m=p_m/p_0$ with $p_0=\sqrt{\hbar m\omega}$. These parameters are related to the variances by $a_{11}=2 {\rm Var (X_m)}=2\sigma^2 (X_m)$ and $a_{22}=2 {\rm Var (P_m)}=2\sigma^2 (P_m)$.  We show the first 20 $\mu$s of the sequence in figure~\ref{zoom_10mK}, allowing one to visualize the oscillations of the plotted parameters. The two curves clearly indicate the alternate increase and decrease of these two parameters. The stroboscopic pulses have been turned on after 2.5 complete oscillations in order to display the initial variance (flat line) of the two variables before applying the squeezing protocol. Figure~\ref{10mK} shows the full 400 $\mu$s duration of the sequence. The oscillations cannot be distinguished and appear as a gray surface on the scale of the figure.

In order to obtain a truly squeezed state in the quantum mechanical sense, the standard deviation $\sigma(X_m)$ or $\sigma(P_m)$ needs to be smaller than $1/\sqrt{2}$ (the Heisenberg uncertainty relation in our dimensionless units states that $\sigma(X_m)\sigma(P_m)\geq 1/2$, and squeezing being defined as one of the variables being lower than its ``fair share'' of uncertainty at the Heisenberg limit, which is equal to $1/\sqrt{2}$). As we plot two times the standard deviation, the plotted quantities must be less than $\sqrt{2}$ in order to claim squeezing. As shown in figure figure~\ref{10mK}, even for long times (400 $\mu$s) the parameters never drop below the solid line for the above-described conditions of a 10 mK bath temperature and a 10 Hz coupling rate to the environment. {\it I.e.}, while the uncertainties are significantly reduced compared to the initial thermal state, quantum squeezing below the width of the ground state fluctuations is not achieved.

\begin{figure}[t]
\centerline{\includegraphics[width=9cm]{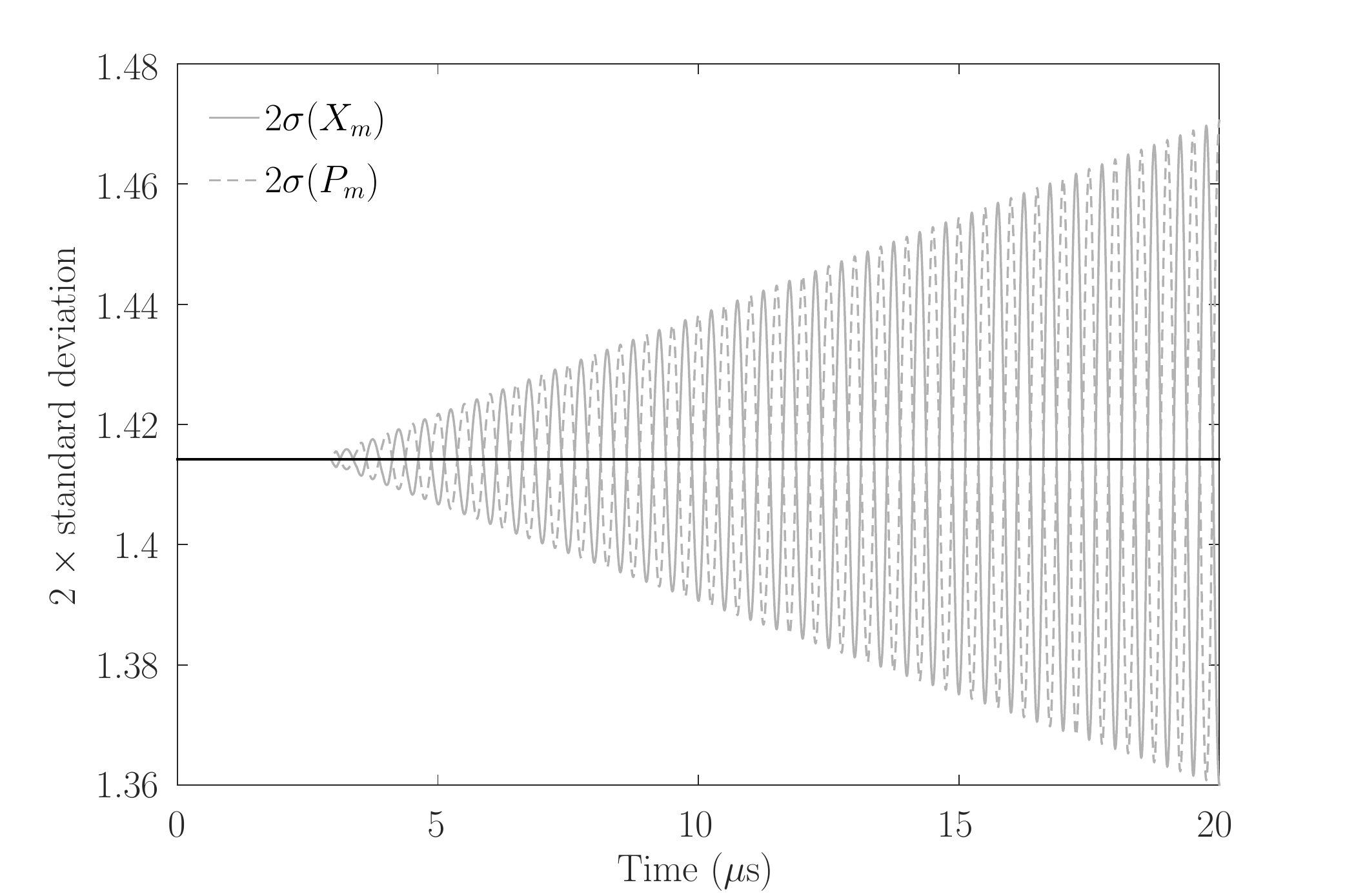}}
\vspace*{8pt}
\caption{The first 20 microseconds of the squeezing sequence with a thermal bath of 0 mK. The initial values of the variances correspond to the Heisenberg limit. As soon as the first pulse of the probe laser is applied, we observe the onset of squeezing.\label{zoom_0K}}
\end{figure}

However, reducing the temperature to 0.7 mK, which is the lowest resonator temperature achievable with our active laser cooling described in previous work \cite{seidelin2019}, and maintaining the coupling rate at 10 Hz, a modest level of squeezing is obtained ($2\sigma$=1.24 at 400 $\mu$s, with $\sigma$ the lowest of $\sigma(X_m)$ and $\sigma(P_m)$. This is shown in figure~\ref{0_7mK}.

Another possibility to reach quantum squeezing is to decrease the coupling rate between the resonator and the thermal bath. Indeed, even with a thermal environment at 10 mK, with $\gamma$=0.1 Hz, a modest effect of squeezing is achieved, with $2\sigma$=1.07 at 400 $\mu$s. This is illustrated in figure \ref{10mk_gamma_0_1}.

\begin{figure}[t]
\centerline{\includegraphics[width=9cm]{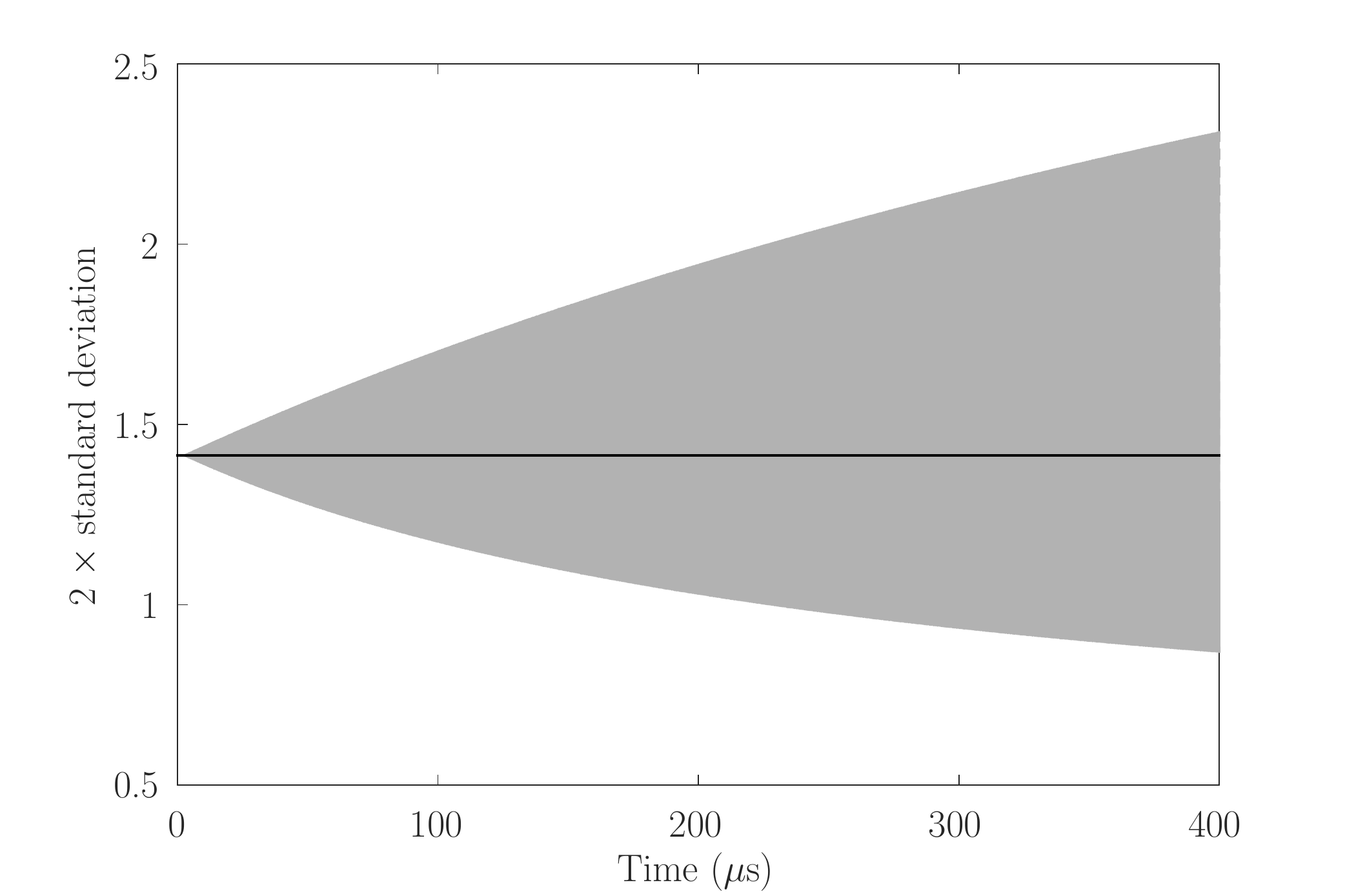}}
\vspace*{8pt}
\caption{The full duration of simulations of squeezed states with an initial resonator temperature of 0 mK shown in figure \ref{zoom_0K}. The squeezing is significant, as illustrated by the gray area which drops well below the solid line ($2\sigma$=0.90 at 400 $\mu$s).\label{zero}}
\end{figure}
In order to achieve more significant squeezing, we need to start out with bath temperature close to zero. In figure~\ref{zoom_0K}, we show the case of an initial temperature of 0 K. In this case, the initial variances are already at the Heisenberg limit, and the squeezing appears as soon as the first pulse is applied. We can, in this case, obtain significant squeezing within 400 $\mu$s ($2\sigma$ reaches 0.90 at 400 $\mu$s), regardless of the coupling rate between the resonator and the thermal bath.\\

It is not experimentally possible to reach a bath temperature sufficiently close to zero ({\it i.e.} having the resonator in the quantum ground state) with current cryogenics technology for the described physical system (50 $\mu$K corresponds to a mean quantum occupation number of the order of 1). Careful design and optimization of the resonator and its clamping loss may however allow reaching ultra high decoupling level from the thermal bath, which seems to be the major conditions that will allow quantum squeezing in the discussed mechanical systems.\\

\section{Entanglement between 2 resonators}

A particularly interesting non-classical state includes 2 entangled resonators, and the condition for achieving such a state can be seen as a direct extension of the condition necessary to squeeze the quadratures of a single resonator. We define the variables

$$X_{\pm}=\frac{1}{\sqrt{2}}(X_{2}\pm X_{1})$$
$$P_{\pm}=\frac{1}{\sqrt{2}}(P_{2}\pm P_{1}),$$

where the subscripts 1 and 2 refers to resonator 1 and 2 respectively. In this notation, the criterion for entanglement of the 2 resonators can be written as\cite{Korppi2018,Duan2000}

$${\rm Var}(X_{+})+{\rm Var}(P_{-})=\sigma^2(X_{+})+\sigma^2(P_{-}) < 1$$

The criterion itself points to a method for preparation of the entangled state, namely by using the dispersive interaction of a optical probe beam with ion ensembles implanted in both resonators such that the optical phase shift accumulates contributions from both $X_1$ and $X_2$, and thus effectively reads out $X_{2} + X_{1}$. The back action of this measurement is precisely the squeezing of the collective observable $X_{+}$, while probing (and squeezing) of the commuting observable $P_-$  leads to the desired entangled state \cite{Duan2000a,Julsgaard2001}.

\section*{Conclusion}
 
We have shown that the system of a strain-coupled rare-earth doped resonator holds promise for exhibiting non-classical states, in particular squeezed states. However, the relatively strong coupling to the environment makes it extremely challenging to obtain squeezed states within a reasonable timescale at a non-zero bath temperature, by using this simple protocol based on stroboscopic pulsing of a probe laser alone. Future work includes optimizing the protocol to allow faster squeezing, in order to decrease the influence of the coupling to the environment, as well as an experimental study of the best approach to increase resonator quality factors, and hence decoupling as much as possible from the thermal bath. Finally, mechanical resonators are potentially useful as highly sensitive force-probes, and future work also includes a study of how squeezed states could benefit the sensitivity of the resonator in this context.

\section*{Acknowledgements}

The project has been supported by the European Union's Horizon 2020 research and innovation program under grant agreement No 712721 (NanOQTech), Ville de Paris Emergence Program, the R\'{e}gion Ile de France DIM C'nano and SIRTEQ, the LABEX Cluster of Excellence FIRST-TF (ANR-10-LABX-48-01) within the Program "Investissements d'Avenir" operated by the French National Research Agency (ANR), and the EMPIR 15SIB03 OC18 program co-financed by the Participating States and from the European Union's Horizon 2020 research and innovation program.

\end{document}